\documentclass[pra,twocolumn,superscriptaddress]{revtex4}

\usepackage{mathrsfs,amsmath,amssymb,graphicx,paralist}

\def\scr{\mathscr}

 \def\SI{{\scr I}}

 \def\SV{{\scr V}}

\def\avg#1{\langle#1\rangle}    \def\<{\langle}         \def\>{\rangle}

\def\al{\alpha}                   

        \def\del{\delta}        \def\Del{\Delta}
\def\eps{\epsilon}

  \def\V0{{\mathbf 0}}

\def\Bx{{\mathbf x}}  \def\B0{{\mathbf 0}}
\def\Br{{\bf r}}   

\def\Bk{{\mathbf k}}  \def\Bq{{\mathbf q}} 
   \def\Bn{{\mathbf n}}
 \def\Be{{\mathbf e}}

     \def\BT{{\mathbf T}}
  \def\BQ{{\mathbf Q}}
\def\BL{{\mathbf L}} \def\BG{{\mathbf G}}
 
\def\Btheta{{\boldsymbol \theta}}

\def\be{\begin{equation}}       \def\ee{\end{equation}}
\def\bea{\begin{eqnarray}}      \def\eea{\end{eqnarray}}
\def\nn{\nonumber} 

\def\Sup{Appendices}

\begin{document} 

\title{Atomic matter of non-zero momentum Bose-Einstein
  condensation and orbital current order}

\author{W. Vincent Liu}
\affiliation{Department of Physics and Astronomy, University of
  Pittsburgh, Pittsburgh, PA 15260} 

\author{Congjun Wu}
\affiliation{Kavli Institute for Theoretical Physics, University of
  California, Santa Barbara, CA 93106}

\begin{abstract}
  The paradigm of Bose-Einstein condensation has been associated with
  zero momentum to which a macroscopic fraction of bosons condense.
  Here we propose a new quantum state where bosonic alkali-metal atoms
  condense at non-zero momenta, defying the paradigm. This becomes
  possible when the atoms are confined in the $p$-orbital Bloch band
  of an optical lattice rather than the usual $s$-orbital. The new
  condensate simultaneously forms an order of transversely staggered
  orbital currents, reminiscent of orbital antiferromagnetism or
  $d$-density wave in correlated electronic systems but different in
  fundamental ways.  We discuss several approaches of preparing atoms
  to the $p$-orbital and propose an ``energy blocking'' mechanism by
  Feshbach resonance to protect them from decaying to the lowest
  $s$-orbital.  Such a model system seems very unique and novel to
  atomic gases. It suggests a new concept of quantum collective
  phenomena of no prior example from solid state materials.
\end{abstract}
\pacs{03.75.Nt,67.40.-w,74.72.-h}

\maketitle


\section{Introduction}

Confining bosonic atoms in an optical lattice can bring out different
and new physics beyond the standard Bose-Einstein condensation (BEC)
observed in a single
trap~\cite{Anderson+Cornell:BEC:95,Davis+Ketterle:BEC:95}.   
The superfluid-Mott insulator experiment
on an optical lattice~\cite{Bloch-SF-Mott:02}, based on an early
theoretical idea
\cite{Fisher:Bose-Hubbard:89,
Jaksch+BruderETAL-ColdBosonicAtomOpticalLattices:98},
demonstrated one such example with bosons.
Proposals of exploring various lattice atomic systems, many concerning
spin, has further extended the scope of interest into different directions,
experimentally~\cite{Paredes:Tonks:04,Weiss:Tonks:04,Stoeferle+Esslinger-Mott1D:04,Browaeys:05,Esslinger:05,Stoferle:06,Kasevich:05pre}
and theoretically (see
Ref.~\cite{Demler-Zhou:02,Hofstetter:02,Kuklov++Duan:03,Kollath:05,Carr:05pre,Zhou:05pre,scarola2005}
and references therein).

Atomic optical lattice not only can realize many standard solid-state
problems but also may bring new and unique aspects specific to the
atomic gas. One possible direction is the orbital physics \cite{Girvin:05}
. This is a
new direction that has not yet received as much attention as
spin.  The
pioneering experiments of  Browaeys {\it et al}~\cite{Browaeys:05} and 
Kohl {\it et al}~\cite{Esslinger:05}, which  demonstrated the
occupation of bosonic and fermionic atoms, respectively, in the higher
orbital bands, further justify and  motivate theoretical
interest in the orbital degrees of freedom of cold atoms beyond the
conventional $s$ orbital band, such as the next
three $p$ orbitals.
In electronic solids such as manganese oxides and other transition metal
oxides, the orbital physics is believed to be essential for
understanding their metal-insulator transitions, superconductivity,
and colossal magnetoresistance.  The solids are of periodic arrays of
ions. The quantum mechanical wavefunction of an electron takes various
shapes when bound to an atomic nucleus by the Coulomb force. For those
oxides, the relevant orbitals of the electron are the $5$ $d$-wave
orbitals (usually split into two groups of $e_g$ and $t_{2g}$ due to
the crystal field).  The orbital degree of freedom, having intrinsic
anisotropy due to various orbital orientations, interplaying with the
spin and charge, gives rise to an arena of interesting new phenomena
in the field of strongly correlated
electrons~\cite{Tokura2000,Maekawa:04bk}.  We will focus
on the orbital degree of freedom of cold atoms below.

In this paper, we point out that the current experimental condition
makes it possible to study a whole new class of lattice system---the
dilute $p$-orbital Bose gas---beyond the conventional $s$-orbital
Bose-Hubbard model.  We show that the system reveals in the superfluid
limit a new state of matter in which atoms undergo Bose-Einstein
condensation at {\it nonzero momenta} and form a staggered orbital
current order simultaneously. 
These features distinguish the $p$-orbital atomic gases from the
$d$-orbital electronic oxide compounds.
The unique signature of
the state is predicted for the time-of-flight experiment.

\section{Atoms in the $p$-orbital}
Let us study an optical lattice of bosonic atoms in a single
internal (hyperfine spin) state. 
To gain a qualitative understanding, we approximate the lattice
potential well by a harmonic potential around minimum.  The
characteristic ``harmonic oscillator'' frequency is $\omega_b =\sqrt{4
V^b_{0} E^b_{R}}$ where $V^b_0$ is the 3D lattice potential depth
and $E^b_R$ the recoil energy for the bosonic atoms.  The recoil
energy is determined by the laser wavelength and atom mass.  In the
presence of periodic lattice potentials, the boson state can be
expanded in the basis of Wannier functions, to be denoted as
$\phi^b_{\Bn}(\Bx)$, with $\Bn$ the Bloch band index.  The lowest
Bloch band is $s$-wave symmetric with $\Bn=(000)$. The next band is a
$p$-wave with three-fold degeneracy, $p_\mu$ with $\mu=x,y,z$,
corresponding to $\Bn=(100),(010),(001)$. The energy splitting between
$s$ and $p$ is $\hbar\omega_b$. [Note that the three degenerate
$p$-wave energy subbands disperse anisotropically when hopping process
included.]
The fact that  the next band starts at the $p$ orbital instead
of the $s$   tells an important difference between 
the optical lattice potential and the (Coulomb) ionic lattice potential in
electronic solids. 

In a dilute weakly interacting atomic boson gas confined in optical lattices, 
bosons intend to aggregate into the lowest $s$ band
in the low temperature limit, with an exponentially small fraction in the
higher Bloch bands, suppressed by the factor $e^{-\hbar\omega_b/k_BT}$.
A single-band approximation is then adequate, which was
proven successful, both theoretically and
experimentally~\cite{Jaksch+BruderETAL-ColdBosonicAtomOpticalLattices:98,Bloch-SF-Mott:02}.

Several approaches are available for transferring cold atoms to the
first excited $p$-orbital band. In the study of a related but
different model, Isacsson and Girvin~\cite{Girvin:05} suggested: (A)
to use an appropriate vibrational $\pi$-pulse with frequency on
resonance with the $s$-$p$ state transition; (B) to apply the method
demonstrated in the experiment of Browaeys et al~\cite{Browaeys:05} by
accelerating atoms in a lattice. We may also add a third possible
approach, that is, (C) to sweep atoms adiabatically across a Feshbach
resonance.  M. Kohl {\it et al}~\cite{Esslinger:05} pioneered this
method experimentally by showing fermionic atoms transferred to higher
bands; the phenomenon was subsequently explained in
theory~\cite{Diener-Ho:06}. Whether bosonic atoms can be transferred 
this way remains to be seen.

Now suppose that a metastable $p$-orbital Bose gas has been prepared
on the optical lattice.
The remaining challenge is that the
system is not in the ground state, and thus
genuinely has a finite life time. 
The interactions between two (bosonic) atoms, although weak, can cause
atoms in the $p$-orbital states to decay. 
An elastic decaying process (which conserves total energy) is that 
two atoms initially in the $p$-orbital band scatter into the final
state of one atom in the
$n=0$ ($1s$) band and another in the $n=2$ orbital ($2s,1d$)
band, 
where $n$ represents the principle (energy
level) quantum number and the states of the harmonic oscillator 
are labeled in the
Landau-Lifshitz notation~\cite{Landau-Lifshitz:77QM}.  For a
related model, Isacsson and Girvin~\cite{Girvin:05} have studied the
decaying rate and estimated that the life time is about $10$ to $100$ times
longer than the time scale of tunneling in an optical lattice.
However, such a life time can be still short to achieve condensation and
perform experimental detection.
In the following, we propose a new mechanism that should 
suppress the above decaying process and thus extend the life time.

\section{Energy blocking of the $p$-orbital decay}

\begin{figure}[htbp]
\includegraphics[width=\linewidth]{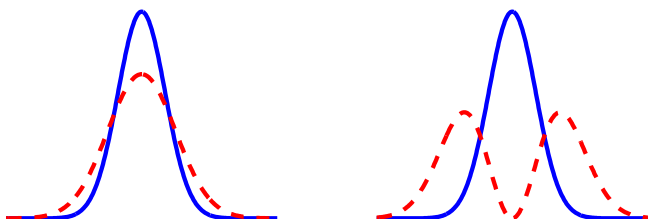}
\caption{(Color online) 
The overlap between $s$ and $p$ density clouds (wavefunctions
  squared) is smaller than between two $s$ clouds.}
\label{fig:sp_overlap}
\end{figure}
\begin{figure}[htbp]
\includegraphics[width=0.9\linewidth]{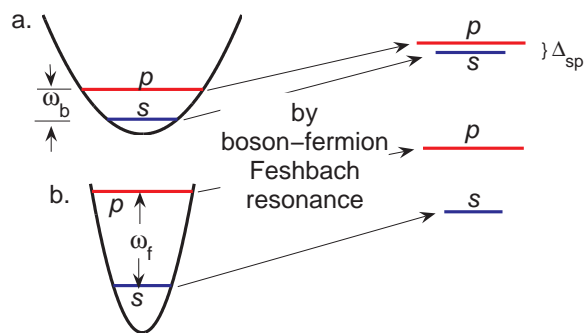}
\caption{(Color online)    Energy band shift 
due to Feshbach resonance between bosons and fermions. 
(a) The bosonic $1s$ band is moved close to the $1p$ band
while the $2s$ and $1d$ bands  (not shown) are both pushed higher.
  (b) The fermions fill up their own lowest $s$-band as an band insulator.
The relative shift of the fermionic band energy is less significant
than that of the bosonic case, for $\omega_f$ can be made far larger than
both $\omega_b$ and the (Feshbach tunable) energy shift.
}
\label{fig:shift_bands}
\end{figure}

We propose a deep fermion optical lattice on top of a relatively shallow 
optical lattice for bosons, such that the characteristic lattice well 
frequencies are very different ($\omega_b\ll \omega_f$). 
Consider to load a fermion density around $n^f\approx 1$ 
(that is, one fermion per site). Now one tunes a
Feshbach resonance
between boson and fermion~\cite{Stan+Ketterle:04,Inouye+Jin:04}, 
to set the inter-species interaction 
strength in energy scale
between $\hbar\omega_b$ and $\hbar\omega_f$.  
In this case, the fermions fill
up the $s$-band completely, so essentially behaving as an band insulator 
whose dynamical
effect on bosons becomes exponentially suppressed by the band
energy gap.  The lowest $s$ orbital wavefunction is approximately a
Gaussian peaked at the center of the lattice well. 
The role of fermions then can be thought as providing a
repulsive central potential barrier, 
in addition to
the optical lattice potential,  for bosons on the same lattice site.
As a result, all energy bands are shifted up
significantly (including $s$ and $p$ of course) by the Feshbach
interaction with the $s$-orbital fermion.
Because the overlap integrals are different as shown in 
Fig. \ref{fig:sp_overlap}, the energy shifts are
different in magnitude for different orbital states
(Fig. \ref{fig:shift_bands}).

The lowest orbit for a single particle in a single site must be nodeless; 
it is actually not possible to increase the $s$-state energy higher
than the $p$-state.
However, the gap $\Del_{sp}$  between ($n=0$) $s$ and ($n=1$) $p$ can be
significantly reduced to very small by a sufficiently strong 
interspecies Feshbach resonance, 
as illustrated in Fig.~\ref{fig:shift_bands}.
In Appendix~\ref{sec:gap_closing}, 
we give an estimate of the interaction strength needed. 
On the other hand, the splittings between the $1p$ state 
and states in the $n=2$ level ($2s$ and $1d$) should not change
much because the wavefunctions of the latter all  are spatially more
extended than the $1s$ state.
In summary, the lowest two Bloch bands ($s$-$p$) are close in energy 
while all other Bloch bands ($n\geq 2$) have energy far above the first
excited $p$-orbital band  (in the energy scale of $\Del_{sp}$).
The low energy quantum theory of the system effectively reduces to a two-band
problem. 

The lattice gas under study is also assumed to be in the tight binding
limit such that the tunneling amplitude is smaller than
$\Del_{sp}$, so much smaller than the band level splitting between the
$p$ and all other higher bands (i.e., $t_\parallel\ll \Del_{sp}\ll
\hbar\omega_b$).  
Any elastic scattering process that scatters atoms
out of the $p$ orbital band must conserve the total energy.  
Under the proper condition described above, 
the decay rate of the $p$-orbital Bose gas must be suppressed
by the energy conservation law: any two atoms initially in the ($n=1$)
$p$-orbital band cannot scatter out of the band by the two-body
scattering process.  [Of course, the scattering within the three
sub-bands of $p$-orbits are allowed.]
This idea is one of our main results  (see
  Appendix~\ref{sec:model:BF} for technical detail).

\section{A lattice gas of $p$-orbital bosons}
The quantum theory of bosonic atoms prepared in the $p$-orbital state
is effectively described by a $p$-band Bose-Hubbard model.  A
standard derivation (see Appendix~\ref{sec:tight-binding})
gives the Hamiltonian
\bea
H&=&\sum_{\Br,{\mu\nu}} [t_\parallel\del_{\mu\nu} -t_\perp(1-\del_{\mu\nu})]
\Big(b^\dag_{{\mu},\Br+a\Be_{\nu}} b_{{\mu}\Br}
+h.c.\Big)\nn\\
&& +{\textstyle {1\over 2}} U \sum_\Br \left[  n^2_\Br -{\textstyle{
      1\over 3}}
 \BL^2_\Br\right]   \label{eq:H:pBH}
\eea
where $a$ is lattice constant.
Here, $b_{\mu\Br},b^\dag_{\mu\Br}$ are annihilation and creation
operators for bosons in  lattice site $\Br$ and orbital state $p_\mu$
(index label $\mu,\nu=x,y,z$); $n$ and $\BL$ are the boson density and angular
momentum operators 
$
n_\Br= \sum_\mu b^\dag_{\mu\Br} b_{\mu\Br}
$, 
$
L_{\mu\Br}=-i\sum_{\nu\lambda} 
\eps_{\mu\nu\lambda} b^\dag_{\nu\Br} b_{\lambda\Br}$.
This model is invariant under U(1) phase
transformation, cubic lattice rotations, and time reversal
transformation.

\begin{figure}[htbp]
\includegraphics[width=0.9\linewidth]{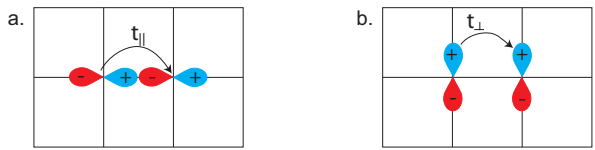}
\caption{(Color online) Anisotropic hopping matrix elements of $p$-orbital bosons
  on a cubic lattice. 
  The longitudinal $t_\parallel$ is in general far greater than
  $t_\perp$ because the overlap integral for the latter is
  exponentially suppressed. The `$\pm$' symbols indicate the sign
  of two lobes of  $p$-orbital
  wavefunction.}
\label{fig:hop}
\end{figure}

The model is determined by the following parameters: 
$t_{\parallel}$ and $t_\perp$ are the nearest-neighbor 
hopping matrix elements
in  longitudinal and transverse directions, respectively, 
with respect to $p$-orbital orientation; and $U$ is the onsite
(repulsive) 
interaction due to the intrinsic (non-resonant) 
$s$-wave scattering between two bosons. 
{In quantum chemistry, $t_{\parallel,\perp}$ are named as
$\sigma(\pi)$-bond, respectively.}
Their precise definition is given in \Sup.  By
definition of the Hamiltonian, both $t_\parallel$ and $t_\perp$ are positive;
note $t_\parallel\gg t_\perp$ for the tunneling overlap is sensitive to
orientation 
(Fig.~\ref{fig:hop}).

The quantum physics of $p$-orbital bosons seems to have never been
studied before except very recently by Isacsson and
Girvin~\cite{Girvin:05} in a different, interesting limit in which the
transverse tunneling $t_\perp$ is completely suppressed.
Different from ours in symmetry and ordering,  their model
had
infinite and subextensive  (local) gauge symmetries and
columnar orderings.  Another difference is that it is not obvious
whether their onsite interaction were SO(3) invariant.

The interaction in the lattice Hamiltonian (\ref{eq:H:pBH}) is
{\it ferro-orbital} ($U>0$), 
suggesting that the bosons at the same site prefer to
occupy the same orbital-polarized state carrying maximal angular
momentum. This is analogous to the
Hund's rule for electrons to fill in a degenerate atomic energy shell
which favors spin-polarized configuration.
Next, we shall
show that the ferro-orbital interaction, together with the $p$-band
hopping, gives rise to an orbital ordered BEC.

\section{Orbital Bose-Einstein condensation (OBEC)}
A weakly interacting Bose gas is expected to
undergo BEC, becoming superfluid at low temperatures. In the optical
lattice, such a state has been firmly established, both experimentally
and theoretically, for bosonic atoms occupying the $s$-band (the
widely studied Bose-Hubbard model). In our $p$-band model when the
interaction is weak and repulsive ($0<U\ll
t_\parallel$), 
the non-interacting term of the Hamiltonian
dominates. 
The $p$-band bosons have the energy dispersion, 
$
\eps_{\mu\Bk}= 2\sum_{\nu} [t_\parallel\del_{\mu\nu}
 -t_\perp (1-\del_{\mu\nu})] \cos(k_\nu a)\,,  
$
where $\mu,\nu$ label the three subbands and $a$ is the lattice constant.

\begin{figure}[htbp]
\includegraphics[width=\linewidth]{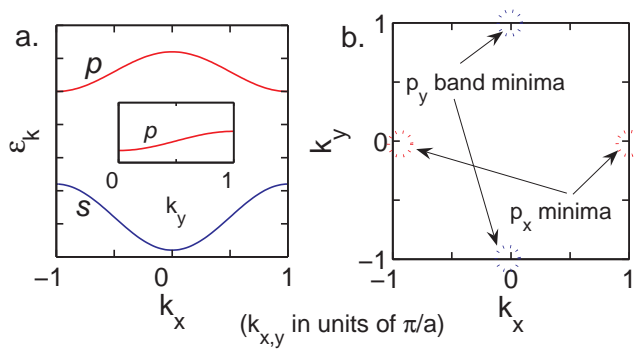}
\caption{(Color online) Illustration of $p$-band dispersions. (a) The energy
  dispersion of the $p_x$ orbital band (as an example) on the $k_x$
  axis in the first Brillouin zone, in comparison with the $s$ band.
  INSET: its dispersion along the line $({\pi\over a},-{\pi\over a}
  ,0)$---$({\pi\over a},{\pi\over a} ,0)$; the energy zero line is
  shifted arbitrarily for display.  Note that the $s$ and $p$
  bands have different locations for minimum energy.  The band gap
  between the two levels is tuned  small
  by the Feshbach resonance.
  (b) The $p$ band minima in
  the $k_x$-$k_y$ plane (the $p_z$'s minima are not shown but can be
  obtained by rotating $90^\circ$ out of the plane).  Orbital BEC
  should occur at a subset of the minima, spontaneously breaking the
  lattice translational and (orbital) rotational symmetry,
  in addition to the U(1).}
\label{fig:p_dispersion}
\end{figure}

There are two new aspects of the $p$-orbital bosons. The first is that BEC
takes place at non-zero momenta. While the paradigm of BEC should
occur at zero momentum, there is no real reason that has
to be so. Only does the lowest energy state matter. 
The $p$-band energy dispersion shows an {\it exceptional} and remarkable
case in which the lowest energy state of boson happens to be at finite
momenta $\BQ_\mu$, defined as
\be \textstyle
\BQ_x=\left({\pi\over a},0,0\right),\  \BQ_y=\left(0,{\pi\over a},0\right),\
\BQ_z=\left(0,0,{\pi\over a}\right)\,,
\ee
for the respective $p$ orbital states (Fig.~\ref{fig:p_dispersion}).  
Note that $\pm\BQ_\mu$ are identified by a reciprocal lattice vector, 
$2 \BQ_\mu=0 (\mbox{mod}\ 2\pi/a)$. 
The condensate at momentum $+\BQ_\mu$ is essentially
  equivalent to that at $-\BQ_\mu$ on the lattice. Therefore, the sum of atoms
  over all modes does not show any net current flow in any direction.
  This feature is important for a possible experimental test of
  the state that we shall elaborate below.

The second aspect is that the orbital degeneracy of
$p$ band bosons opens possibilities of novel orbital physics.
Orbital ordering, typically 
involving $d$-orbital (fermionic)
electrons that are argued to be essential
to understanding a class of strongly correlated 
transition-metal oxides, has become a topical subject
in condensed matter physics~\cite{Tokura2000}. 
In our $p$-band model, condensing into any of
these momenta  $Q_\mu$ or their linear superpositions
equally  minimizes the total kinetic energy.
It thus provides a first bosonic version of such kind 
from atomic physics.
Being ferro-orbital, the interaction in the Hamiltonian will be shown
later to favor a
condensate of angular momentum ordering, to have
$\avg{\BL_{\Br}^2}\neq 0$.

Keeping the above  aspects in mind, we seek a condensate 
described by the following order parameter of total
six real variables,
\be
\begin{pmatrix}
\avg{b_{x\Bk=\BQ_x}}\\
\avg{b_{y\Bk=\BQ_y}}\\
\avg{b_{z\Bk=\BQ_z}}
\end{pmatrix}
=\rho e^{i\varphi-i\BT\cdot{\Btheta}}
\begin{pmatrix}  \cos\chi \\ i\sin\chi
 \\ 0\end{pmatrix}\,,
\label{eq:<b>=SO3}
\ee
with $\BT=(T_x,T_y,T_z)$ 
the generators of $SO(3)$ orbital rotation in the
following matrix representation (an adjoint ${\bf
  3}$ for the group theory experts) defined by its elements:
$
[T_\mu]_{\nu\lambda}=-i{\eps}_{\mu\nu\lambda}\,, \quad 
\mu,\nu,\lambda=x,y,z\,.
$
Such parameterization  manifests symmetry: 
$\varphi$ is the overall phase of the U(1) symmetry; the three angle
variables, $\Btheta=(\theta_x,\theta_y,\theta_z)$, are the orbital
rotation of SO(3) and $\chi$ changes sign under time 
reversal; $\rho$ is the modulus field, fixed by the total boson
density in the condensate $n^b_0$ through $\rho=\sqrt{\SV n_0^b}$
where $\SV$ is the  3D lattice volume in the units of
$a^3$ (so $\SV$ dimensionless).  

We now proceed to calculate the mean field interaction energy.  While
the full Hamiltonian only has the cubic lattice symmetry, the
interaction term, being onsite, enjoys however  a continuous SO(3)
rotation invariance. Therefore, the symmetry dictates that the mean
value of the interaction term be independent of $\varphi$ and $\Btheta$.
(The former is due to the exact U(1).)  This argument reduces our
calculation essentially to a problem for a single-variable order
parameter, 
namely, $\chi$. 
Then, a straightforward evaluation of the mean value of the
interaction energy determines
$$
{\avg{H_{U\,\mathrm{term}}}} = 
{\textstyle {1\over 2}} {\SV U  {n^b_0}^2}\left[1-{\textstyle {1\over
      3}}  \sin^2(2\chi)\right]\,.
$$
The interaction energy is minimized at
$
\chi=\chi_\pm\equiv \pm {\pi\over 4}\,.
$

The two minima, $\chi=\chi_\pm$, are  degenerate and discrete,
reflecting 
the time-reversal symmetry. Of course, shifting the value of
$\chi_\pm$ by $\pi$ gives other minima of the same energy, but the
$\pi$ shift can be absorbed away by an overall U(1) phase adjustment
or an orbital rotation. 
The order parameter points to one
of the two discrete minima, 
spontaneously breaking the  time
reversal symmetry.
The resultant 
quantum state is an axial superfluid having a macroscopic
angular momentum ordering. The order parameter is then found to be
(say for $\chi=\chi_+=+{\pi\over 4}$)
\be
\begin{pmatrix}
\avg{b_{x\Bk=\BQ_x}}\\
\avg{b_{y\Bk=\BQ_y}}\\
\avg{b_{z\Bk=\BQ_z}}
\end{pmatrix}
=\sqrt{\SV n^b_0\over 2}
\begin{pmatrix}  1 \\ i
 \\ 0\end{pmatrix}  \label{eq:<b>=xy}
\ee
with a degenerate manifold characterized by
phase $\varphi$ and rotational angles
$\Btheta$. This is our $p$-orbital BEC ($p$-OBEC). It breaks the U(1) phase,
lattice translation, orbital SO(3) rotation 
and time reversal symmetries. Quantum or
thermal fluctuations, which will be studied in the future, are
expected to break the SO(3) rotational symmetry and
align the orbital condensate to specific lattice
directions.

\section{Staggered orbital current}
The $p$-OBEC contains one novel feature that is absent in the
conventional BEC. To illustrate this, we assume that the lattice has a
small anisotropy such that the state is pinned to the $xy$ plane, i.e., the
$p_x\pm ip_y$ order (see the phase diagram in Appendix~\ref{sec:aniso}). 
The novel feature is contained in the structure
factor of the boson number and angular momentum operators.  Taking for
example the state of $\chi=\chi_+={\pi\over 4}$, we found
$
\avg{\BL_{\Bq}} =\left(0,\, 0\,, {n^b_0} 
 \del_{\Bq,\BQ_x+\BQ_y}\right)\,.
$
The momentum dependence of the angular momentum operator reveals that
the $p$-OBEC is also an orbital current wave, in analogy with a
commensurate spin density wave order in antiferromagnets. In real
space, the  order has the {\it staggering} pattern:
$
\avg{L_{x\Br}}=
\avg{L_{y\Br}}=0,
\avg{L_{z\Br}}=n^b_0 (-)^{x+y\over a}\,. 
$
We shall call it  transversely staggered orbital current (TSOC)
for the direction of
$L_z$ alternates only in $x,y$-directions. 
The reader should bear in mind that at mean field level 
the direction of orbital ordering  is arbitrary for a fully SO(3)
invariant interaction.
In real experiments, 
the presence of a symmetry-breaking perturbation, such as a
weak anisotropy in lattice potentials we assumed, or the effect of 
quantum fluctuations  
is expected to pin down the direction.

\begin{figure}[htbp]
\includegraphics[width=0.8\linewidth]{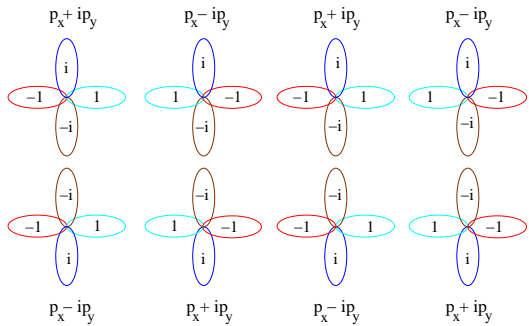}
\caption{(Color online) The real space configuration of the TSOC state, which
exhibits a staggered and uniform  orbital current pattern 
in the $xy$ plane and along the $z$-axis, respectively.
}
\label{fig:pobec}
\end{figure}

We next provide an explanation for the appearance of TSOC 
in real space (Fig.~\ref{fig:pobec}).
The interaction favors a maximum angular momentum at each site. So
the orbit configuration on each site is $p_x\pm ip_y$, corresponding
to angular momentum quanta $\pm\hbar$ per atom. 
On the other hand, the longitudinal and transverse 
hopping amplitudes are of opposite sign.
In order to  maximally facilitate the inter-site hopping,
the phases of $p_{x,y}$ orbits  
should be staggered in the longitudinal direction, and uniform in 
the transverse directions.
As a result, $\avg{L_{z\Br}}$ or the orbital current exhibits a 
staggered (uniform) pattern in the $x$-$y$ ($z$) directions.
This state bears some similarity to its fermionic counterpart of
the orbital antiferromagnetism or $d$-density
wave (DDW)~\cite{CHAKRAVARTY2001} proposed for the high temperature
cuprates.
A major difference is that the DDW current flows on bonds around each
plaquette  with staggered magnetic moments through lattice
whereas the current here circulates inside each site.

\section{Experimental signature}
In a time-of-flight experiment which has been widely used to probe the
momentum distribution of cold atoms, the $p$-OBEC will distinguish
itself from a conventional $s$-BEC with unique structural
factor.
\begin{figure}[htbp]
\includegraphics[width=0.8\linewidth]{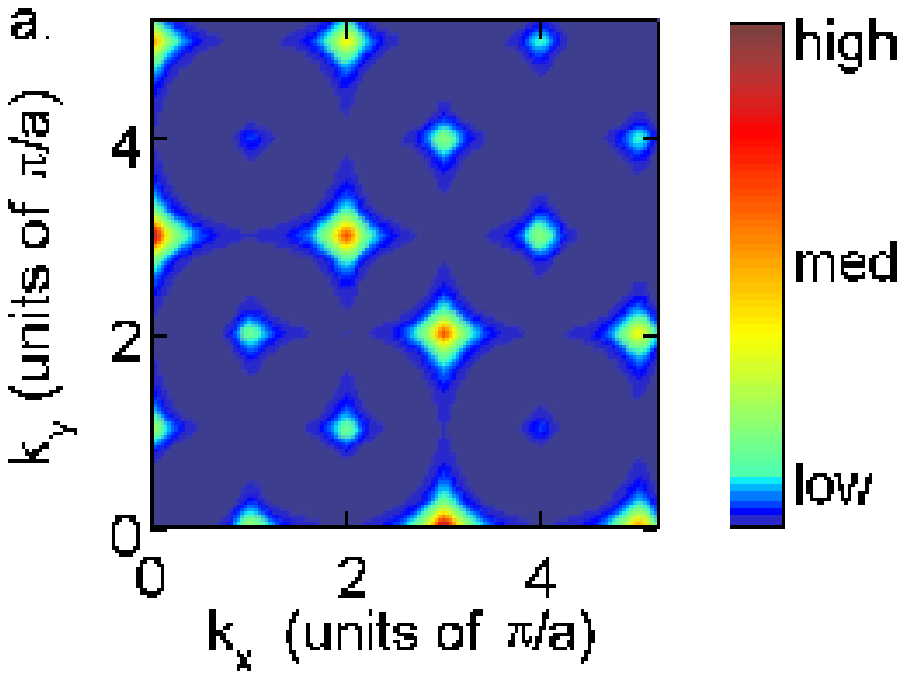}
\includegraphics[width=0.8\linewidth]{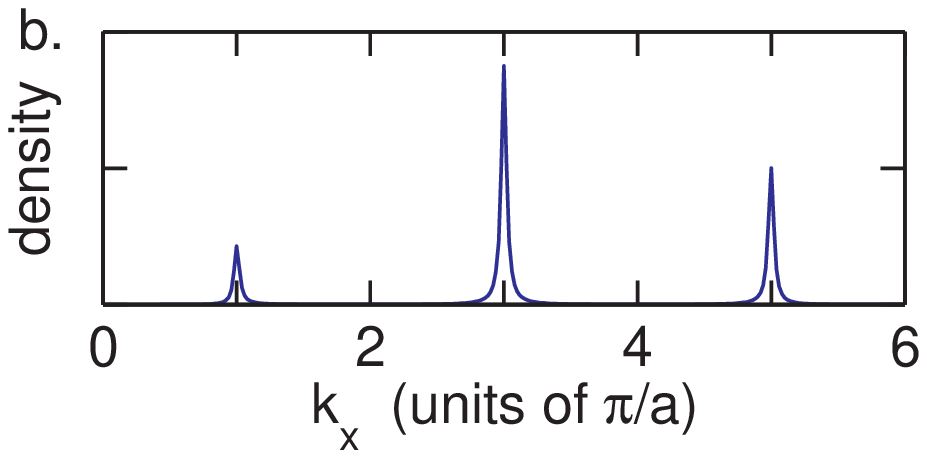}
\caption{(Color online) Prediction of density of atoms for time-of-flight
  experiment. (a) Density integrated over $z$-axis is shown in a
  quarter of $k_x$-$k_y$ momentum plane, with other quarters obtained by
  reflection symmetry. (b) Density shown along the
  $k_x$-axis. Assuming a free-expansion, the atom density at distance
  $\Br$ is $\avg{n^b(\Br)}_\tau\propto
  |\phi^b_{p_x}(\Bk)|^2 \del^3(\Bk-\BG-\BQ_x) +
  |\phi^b_{p_y}(\Bk)|^2 \del^3(\Bk-\BG-\BQ_y)$ where $\Bk ={m\Br/
  \hbar\tau}$, $\tau$ is the time of flight, and $\BG$ runs over all
  three dimensional reciprocal lattice vectors. The absence of peak at
  $\Bk=0$ distinguishes the $p$-OBEC from paradigmatic BEC.  The
  highest peak is not necessarily the closest to zero momentum origin,
  a unique feature of orbital BEC, depending on the size $l_b$ of the
  bosonic Wannier function
  $\phi^b_{\mu}(\Br)$. Parameters are: $l_b/a=0.1$; the $\delta$-function
  is replaced by a Lorentzian line for display. }
\label{fig:tof}
\end{figure}
There are two new aspects (Fig.~\ref{fig:tof}). 
The first is that the
condensation peaks are not located at zero momentum, nor at any other
momenta related to zero by a reciprocal lattice vector $\BG$.  The
second is that, unlike the $s$-wave case, the $p$-wave Wannier function
superposes a {\it non-trivial profile} on the height of density peaks. 
As a result, the highest peaks are shifted
from the origin---a standard for the $s$-wave peak---to the
reciprocal lattice vectors 
whose magnitude is around $1/l_b$. In the following, we show  the
detail of calculations that lead to the above predictions.

We consider the density distribution of the time-of-flight experiment,
assuming ballistic expansion. 
In our $p$-orbital case, it can be written as
\bea
\avg{n(\Br)}_t =\left(\frac{m}{\hbar t}\right)^3 \sum_{\mu,\nu}
\phi^*_{p_\mu} (\Bk) \phi_{p_\nu} (\Bk) \avg {b^\dagger_\mu (\underline\Bk)
b_\nu (\underline\Bk)},
\eea
where $\mu,\nu=x,y,z$; $\Bk=m \Br/(\hbar t)$; $\phi_{p_\mu}(\Bk)$ is the
Fourier transform of the $p$-orbital band Wannier function
$\phi_{p_\mu}(\Br)$; 
and $\underline \Bk= \Bk$ mod { (primitive reciprocal
  lattice vectors)}.
Taking into account that bosons condense into the $p_x+ip_y$ orbital state
spontaneously (according to Eq.~\ref{eq:<b>=xy}),
we arrive at
\bea
\avg{n(\Br)}_t &\propto&
\sum_{\BG}
\Big\{ |\phi_{p_x}(\Bk)|^2 \delta^3 (\Bk-\BQ_x-\BG)\nonumber\\
&+& |\phi_{p_y}(\Bk)|^2    \delta^3 (\Bk-\BQ_y-\BG)
\Big \}.
\eea
That means that 
the Bragg peaks are at $ (\frac{(2m+1)\pi}{a}, \frac{2n \pi}{a}, 
\frac{2l\pi}{a})$ and $(\frac{2m \pi}{a}, \frac{(2n+1)\pi}{a}, 
\frac{2l \pi}{a})$ where $m,n,l$ are integer numbers. 

The Fourier transform of the $p$-wave Wannier orbits exhibits the
following non-trivial form factor,
\bea
|\phi_{p_x}(\Bk)|^2&= & |\phi_1(k_x) \phi_0(k_y) \phi_0 (k_z)|^2,\nn \\
|\phi_1(k)|^2&\propto& (k l_b)^2 e^{-(kl_b)^2/2},  \\
|\phi_0(k)|^2 &\propto &  e^{-(kl_b)^2/2}\,,\nn 
\eea
where $\phi_{0,1}$ refer to the ground and the first excited state
of the  one-dimensional harmonic oscillator, respectively,  and $l_b$
is the oscillator length of the boson optical lattice
potential. 
As a result, unlike the conventional 
$s$-orbital case where the highest weight is
located at the origin of the reciprocal lattice with a distribution width about
$1/l_b$, the $p$-orbital case has the highest weight shifted
from the origin to the reciprocal lattice vectors around $1/l_b$.
In Fig. \ref{fig:tof}, where we used the parameter 
$1/l_b =\frac{10}{a} \approx \frac{3\pi}{a}$, 
the highest intensity thus appears
at the second Bragg peak.

\section{Conclusion and Discussion} In conclusion, we have proposed
a new state of matter in which $p$-orbital bosonic atoms condense at
separate non-zero momenta according to orbital orientations, defying
the zero momentum hallmark characteristic of all standard
Bose-Einstein condensates.  This new state is a prediction for a
$p$-orbital Bose gas on lattice.  We have shown how bosons, in Feshbach
resonance with fermions, can be effectively blocked from occupying the
$s$ band by the energy conservation.  Our idea may be generalizable to
a mixture of two species of bosons.
Realizing this new state ($p$-OBEC) will change the standard way of
thinking BEC as normally attached to zero momentum.

The topic of ultracold atomic gases has been flourishing at the
interface between atomic and condensed matter physics.  The latest
progress in Feshbach resonance and optical lattice has further
extended its scope of interest. One current focus in correlated 
quantum condensed matter is the spontaneous time-reversal symmetry
breaking ground states, for example, the $d$-density wave state
\cite{CHAKRAVARTY2001} proposed as a competing order for the pseudogap
phase of the high T$_c$ superconductivity and the incommensurate
staggered orbital current phase suggested as the mechanism for the
hidden order transition in the heavy fermion system
URu$_2$Si$_2$~\cite{chandra2002}.  Unfortunately, the experimental
observation of these states so far remains elusive and controversial.
The $p$-OBEC state we proposed here is perhaps the first bosonic
example of this kind from ultracold atomic gases.  
The extraordinary
controllability of the atomic system, widely recognized by far, opens
up the possibility of observing this kind of novel states for the
first time.

\section*{Acknowledgment}
We acknowledge the Aspen Center for Physics where this work was
initiated during the Workshop on Ultracold Atomic
Gases. W. V. L. is supported in part by  ORAU through the 
Ralph E. Powe Junior Faculty Enhancement Award.   
C. W. is supported by the NSF
under Grant No. PHY99-07949.

{\it Note Added.}  
Upon the completion of this manuscript,
there appeared an independent work~\cite{Kuklov:06pre} that also
proposes a $p$-orbital BEC based on a related but different mechanism.


\appendix

\section{The microscopic model for a Bose-Fermi mixture}
\label{sec:model:BF}

Our model 
system is a gas of two species of atoms, one being bosonic and another
fermionic, 
confined in two overlapping sublattices with separate potential heights.
The Hamiltonian is 
\bea
H&=& \int d^3\Bx \Big\{\sum_{\al=b,f}
\psi^\dag_\al\big[ {\textstyle -{\nabla^2\over 2m_\al}}
+V_\al(\Bx)\big] \psi_\al\nn \\
&& + g_\mathrm{res} \psi^\dag_f\psi^\dag_b \psi_b\psi_f 
+ g \psi^\dag_b\psi^\dag_b \psi_b\psi_b \Big\}  \label{eq:H_BF}
\eea
where the indexes $\al=b,f$ label the boson and fermion species,
respectively, $g_\mathrm{res}$ is the inter-species interaction tuned by a
Feshbach resonance, and $g$ is a weak
repulsive interaction between bosons themselves. The
(single-component) fermions do not interact between themselves at short
range.
$V_{b,f}(\Bx)$ are the 3D optical lattice potentials constructed 
by counter propagating laser beams. We assume
\be
V_\al(\Bx) = V_{\al 0} \sum_{\mu=1}^3 \sin^2(k_L x_\mu) \,, \quad \al=b,f
\ee
with $k_L$ the wavevector of the light. 
The recoil energy for each species is 
$E^\al_R=\hbar^2 k_L^2/(2m_\al)$, assumed 
different between boson and fermion for different masses.  
In the presence of such periodic 
potentials, the boson and fermion operators $\psi_\al(\Bx)$ can be
expanded in the basis of Wannier functions, $\phi^\al_{\Bn}(\Bx)$,
with $\Bn$ the band index.  Including the lowest and first excited Bloch bands 
($s$-wave and $p_\mu$-wave), we write 
\bea
\psi_b(\Bx) &=& \sum_\Br \Big[b_{s\Br} \phi^b_s(\Bx-\Br) 
 +\sum_\mu b_{\mu\Br} \phi^b_{p_\mu}(\Bx-\Br)\Big] \,, \label{eq:psi->b} \\
\psi_f(\Bx) &=& \sum_\Br \Big[f_{s\Br} \phi^f_s(\Bx-\Br) 
 +\sum_\mu f_{\mu\Br} \phi^f_{p_\mu}(\Bx-\Br) \Big]\,, \label{eq:psi->f}
\eea 
where $\mu=x,y,z$ label the three $p$-orbital bands.  In the harmonic
approximation, the $s$ and $p$ orbital states are directly given by
the harmonic oscillator  (HO) eigenfunctions,
$\phi^\al_{\Bn}(\Bx) = 
[\phi^\al_{n_x}(x)\,
\phi^\al_{n_y}(y)\, \phi^\al_{n_z}(z)]_\mathrm{HO}$, with
$\Bn=(000)$ for $s$-band and $\Bn=(100),(010)$ and $(001)$ for $p_x$, $p_y$
and $p_z$, respectively, in Cartesian coordinates. The basis
functions are kept separate between fermion and boson for they can have
different lattice potential depths and atomic masses.

\subsection{Hartree approximation of
  interspecies interaction}

We examine possible configurations of the band occupation that may minimize 
energy. 
There are several:
both fermions
and bosons in the $s$-band (ss); 
fermions in the $s$-band and bosons in the $p$-band (FsBp); and 
fermions in the $p$-band and bosons in the $s$-band (FpBs). 
The three configurations  have different
interspecies interaction energies per boson-fermion pair as
follows:
\bea
I_{ss}  &=& g_\mathrm{res} \SI_{(000);(000)}\nn \\
& =& {g_\mathrm{res}\over
  [(l_b^2+l_f^2)\pi]^{3/2}}\equiv W \,,  \\
 I_\mathrm{FsBp}&=& g_\mathrm{res} \SI_{(000);(100)}= {W l_f^2\over
  l_b^2+l_f^2} \equiv W_{\times}  \,,  \\
I_\mathrm{FpBs} &=& g_\mathrm{res} \SI_{(100);(000)}= {W l_b^2\over
  l_b^2+l_f^2}\equiv W_{\times}^\prime  \,, 
\eea 
where 
$\SI_{\Bn;\Bn'}\equiv\int d^3\Bx |\phi^f_{\Bn}(\Bx)|^2
|\phi^b_{\Bn'}(\Bx)|^2$, and 
$l_{b,f}$ are the harmonic oscillator lengths for boson and
fermion, respectively, $l_\al \equiv \sqrt{\hbar/(m_\al\omega_\al)}$.

\subsection{Condition for band gap closing}
\label{sec:gap_closing}
To further achieve a simpler effective model, let us  examine
the single-particle 
Hamiltonian at a single site, say at $\Br$,  in turn for boson and fermion.
We shall show quantitatively how one species shifts the energy levels of
another, through the Feshbach interaction. First, consider the
effects of an $s$-band of fermions  on 
bosons.  At Hartree approximation, the single-particle, onsite
energy is shifted due to the interspecies (Feshbach) interaction with
a term in Hamiltonian
\bea
H^b_\mathrm{single \ site} 
&=& \left({\textstyle {3\over 2}}\omega_b +
Wn^f\right) b^\dag_{s\Br} b_{s\Br} \nn \\
&& \hspace{0.1em}
+ \sum_\mu \left({\textstyle {5\over 2}}\omega_b +
W_{\times}n^f\right) b^\dag_{\mu\Br} b_{\mu\Br} \,, \label{eq:H^b_00}
\eea
where $n^f$ is the number of fermions per site, assumed all in the
$s$-band. Likewise, an $s$ band of bosons shift up both the $s$ and $p$
band energies of fermion, yielding a similar term in Hamiltonian,
\bea
H^f_\mathrm{single\ site} 
&=&H^b_\mathrm{single\ site}[b\rightleftharpoons f;
W_\times\rightarrow W'_\times]
\eea
and $n^b$ is the number of bosons per lattice site, all in the $s$. 
For $\omega_f\gg \omega_b$, the energy shift for the fermionic bands is
small compared with the band energy spacing $\hbar \omega_f$; the $s$
band remains lower than the $p$. 
The situation is opposite for boson in this Hatree treatment. 
The ordering of energies of the $s$ and $p$ bands, if one naively
trusts the Hartree approximation,  can be even 
reversed. 

Here we provide a Hartree estimate for 
how strong the interspecies Feshbach resonance is needed
to close the $s$-$p$ band gap for bosons.
The condition for the effect can be met by requiring that the $s$ band  Hatree
energy be higher than that of the $p$, 
\be
\omega_b + W_\times n^f < W n^f\,
\ee 
from Eq.~(\ref{eq:H^b_00}).  
This condition is satisfied when the
Feshbach resonance scattering length $a_\mathrm{res}$, related to
$g_\mathrm{res}$ by $g_\mathrm{res}= {2\pi a_\mathrm{res} (m_b+m_f)
  /(m_b m_f)}$,  is sufficient large such that
\be
a_\mathrm{res}  > {\sqrt{\pi}\over 2}\frac{1}{n^f} \frac{m_f}{m_b+m_f}
\left(\frac{l_b^2+l_f^2}{l_b^2}\right)^{5\over 2}\, l_b \equiv
a^\mathrm{min}_\mathrm{res} \,. \label{eq:a_F>}
\ee
[The derivation implicitly assumed $n^f\leq 1$ to avoid higher band
  complication due to Pauli exclusion.]
This condition  can be achieved
without tuning the scattering length $a_\mathrm{res}$ to very large
via Feshbach resonance, if the minimally required $a_\mathrm{res}$ 
is made sufficiently
small. That can be done by tuning the depth of optical potentials to
make an adequately small $l_b$ and keep $l_f\ll l_b$ in the same time.
As we will discuss next, our condition for band
gap closing 
can in principle avoid the three-body loss problem in the experiments,
by means of operating the system adequately far from the resonance point.

In the experiment of resonant atomic gases, there is always the
issue of finite life time.
Near the Feshbach resonance, a weakly bound dimer relaxes into deep
bound states after colliding with a third atom.  The released binding
energy transformed into the kinetic energy of atoms in the outgoing
scattering channel that will escape from the trap.  Such a process of
three body collision determines the life time of the trapped gas.  The
relaxation rate is believed to be highest in the boson-boson resonance
and lowest in the fermion-fermion resonance, with the rate for the
boson-fermion resonance in between~\cite{Petrov+:04}---though, there
is no explicit calculation for the boson-fermion resonance yet to the
best of our knowledge.  More relevantly, two recent experiments have
observed the boson-fermion interspecies Feshbach resonances in the
systems of $^6$Li-$^{23}$Na
atoms~\cite{Stan+Ketterle:04} and $^{40}$K-$^{87}$Rb
atoms~\cite{Inouye+Jin:04}, respectively, which seem to be very 
stable.  This is part of the reason
that we are
proposing a Bose-Fermi mixture as opposed to a generally less stable
single-statistics Bose gas.    [Another important reason is of course that
the fermion component of the mixture can be made be a band insulator with
virtually no dynamical effects on bosons other than providing a
central potential barrier; see the main text.]

On general ground, we expect that the larger the detuning from the
resonance is, the smaller the three-body loss rate should be, obeying
some power law suppression.  In our model, the estimated scattering
length between fermion and boson is of the order of $l_b
m_f/(m_b+m_f)$ (see above) to enable the effect of $s$-$p$ band gap
closing for bosons.  It can be made sufficient small.  For example,
this can be as small as $50$ Bohr radius for a $^{23}$Na-$^6$Li
mixture, taking lattice constant $a\simeq 400$nm and the $^{23}$Na
oscillator length $l_b\simeq 0.1 a$. That means that the system does
not have to be operated close to the resonance. This makes the
relaxation lifetime practically infinitely long, so the three-body
loss problem is experimentally avoidable.

Finally, in contact with the discussion on the energy blocking of the
$p$ orbital decay in the main text, we do not require that the $s$-$p$
band gap $\Del_{sp}$ be vanishing but just small compared with that
between the $p$ and higher bands.  Therefore, the above quantitative
evaluation of 
the resonance interaction strength is but an estimate for
the scale.

\subsection{Is an equilibrium nodal Bose-Einstein condensate possible?}
 
The mean-field Hartree argument would imply that when the
boson-fermion interaction is tuned strong enough
($a_\mathrm{res}>a^\mathrm{min}_\mathrm{res}$), the $p$ orbital state
has lower energy than the $s$ (see Fig.~\ref{fig:shift_bands}).
However, if all the effect of fermions on boson is replaced by a
single-particle central potential barrier at every lattice site as in
the Hartree approximation, the lowest orbit of the single-boson state
must be nodeless (which is $s$-wave but likely extended) from simple
quantum mechanics.  The single-particle Hartree 
argument must be wrong  in predicting 
the $p$ orbital be a ground state. For a bosonic many-body
system (such as $^4$He liquid), a similar conclusion is
expected,  due to Feynman, who drew an analogy between a single particle and a
many-body system and argued that the $^4$He ground state wavefunction
be nodeless.

Can the $p$-orbital Bose-Einstein condensate be a true equilibrium
(ground) state instead of a metastable state as we described so far?  We
believe the possibility is not completely ruled out by Feynman's
argument. Let us explain it.  Feynman did not specify the requirement
of the form of the two-body interaction but a careful examination of
his argument would show that the whole argument implicitly assumes a
short-range interaction compared with the average inter-particle
distance.  For a finite or long range interaction, smoothing out the
wavefunction will definite lower the kinetic energy per each particle
coordinate but in the same time will necessarily increase the
interaction potential of all neighbor particles within the scope of the
interaction range. In other words, the energy cost or saving is about
the completion of one particle kinetic energy versus many-particle
interaction energy. When the interaction range is long enough, we
conjecture that the many-body effect becomes dominant.  Another way to
think of our argument is that a long-range interaction has a strong
momentum dependence after Fourier transformation.  We can always think
of a potential that is expandable about zero or some characteristic
momentum. That corresponds to derivative terms of the real-space
interaction potential. Once the two-body interaction potential explicitly
involves derivative terms, Feynman's argument seems to fail.

A long-range interaction is not exotic for cold atoms. For atoms in a
{\it narrow} Feshbach resonance, it is known that the effective
interaction between two atoms can be longer than or comparable with
the average inter-particle distance.~\cite{Petrov:KITP:04tk}
  Given the rapid
advancement in the control of atomic gases, it seems not entirely
impossible to realize an equilibrium, not just metastable, nodal 
Bose condensate
in the future.

\section{Tight-binding approximation for 
the $p$-band Bose-Hubbard model}
\label{sec:tight-binding}
In this section, we use the tight-binding approximation to derive the
general $p$-band Bose-Hubbard Hamiltonian where each lattice site 
is approximated as a three-dimensional anisotropic
harmonic potential with frequencies
$ \omega_{b\mu} ~(\mu=x,y,z)$ in three directions respectively.

The free boson Hamiltonian includes the 
hopping and onsite zero-point energies, i.e.,
\bea
H_0&=&\sum_{\Br{\mu}} t_{\mu;\nu}
\Big(b^\dag_{{\mu},\Br+a\Be_{\nu}} b_{{\mu}\Br}
+h.c.\Big)
+ \sum_{\Br\mu} \hbar \omega_\mu  b^\dag_{{\mu}\Br} b_{{\mu}\Br},
\nn \\
\eea
where the hopping amplitudes are determined by
\bea
t_{\mu;\nu} &=& \int d^3 \Bx {\phi^b_{p_\mu}(\Bx)}^*[-{\nabla^2\over
  2m_b}+V_b(\Bx)] \phi^b_{p_\mu}(\Bx+a\Be_\nu) \nn \\
&=& t_\parallel \del_{\mu\nu} -
(1-\del_{\mu\nu}) t_\perp.
\eea
In this definition, both $t_\parallel$ and $t_\perp$ are positive and
$t_\parallel\gg t_\perp$ in general (Fig.~\ref{fig:hop}).
Here we have neglected the difference among the values of
$t_\parallel$ and $t_\perp$ in $x,y,z$ directions
because the major anisotropic effect comes from the
onsite energy difference among $\omega_{x,y,z}$.

In momentum space, $H_0$ reads
\be
H_\mathrm{hop}=  \sum_{\Bk\mu} (\eps_{\mu\Bk} +\hbar \omega_\mu)
b^\dag_{\mu\Bk}b_{\mu\Bk} 
\ee 
with the energy dispersion of boson 
$
\eps_{\mu\Bk}= 2\sum_{\nu} [t_\parallel\del_{\mu\nu}
 -t_\perp (1-\del_{\mu\nu})] \cos(k_\nu a) \,
$
($a$ is the lattice constant.).
The momentum representation of $H_0$ is useful in determining how the $p$ band
bosons condense.

The short-range interaction between bosonic atoms 
in the original microscopic model gives rise to an  
on-site interaction energy between bosons in the $p$ orbitals. It 
can be classified into three terms,
\bea
H_{int1}&=&\frac{1}{2} \sum_{\Br,\mu} U_\mu n_{\Br\mu}  (n_{\Br\mu} -1), \nn \\
H_{int2}&=& \sum_{\Br,\mu\neq\nu} V_{\mu\nu} n_{\Br\mu} n_{\Br\nu}, \nn \\
H_{int3}&=& \frac{1}{2} \sum_{\Br\mu\neq\nu} V_{\mu\nu}
p^\dagger_{\Br\mu}  p^\dagger_{\Br\mu}  p_{\Br\nu}  p_{\Br\nu},
\eea
with
\bea
U_\mu& =& g\int d^3\Bx |\phi_{p_\mu}(\Bx)|^4 \,,  \nn \\
V_{\mu\nu}&=& g\int d^3 \Bx |\phi_{p_\mu}(\Bx)|^2 
|\phi_{p_\nu}(\Bx)|^2 .
\eea
By straightforward calculation, we obtain the following relations
regardless of the anisotropy of the lattice potential,
\bea
U&=&U_{x}=U_{y}=U_z,  \ \ \  V=V_{xy}=V_{yz}=V_{zx}, \nn \\
U&=& 3 V=  \frac{3g}{4 (2\pi)^{3/2}\, l_x l_y l_z },
\eea
where $l_{b}=\sqrt{\frac{\hbar}{m\omega_b}}$ are the harmonic
oscillator 
length. 

As a result, the interaction part can still be reorganized as in Eq.
\ref{eq:H:pBH} as
\bea
H_{int}=\frac{U}{2} \sum_{\Br} [n_\Br^2-\frac{1}{3} \BL_\Br^2].
\label{eq:anisint}
\eea
Note that $H_{int}$ is surprisingly 
the same as that in the isotropic case of Eq.~\ref{eq:H:pBH},
even if the lattice potentials are anisotropic. 
After setting $\omega_{x,y,z}=\omega_b$, the general 
Hamiltonian in above reduces to Eq. \ref{eq:H:pBH} for the isotropic case.

\section{The case of an anisotropic lattice potential}
\label{sec:aniso}

\begin{figure}[htbp]
\includegraphics[width=0.8\linewidth]{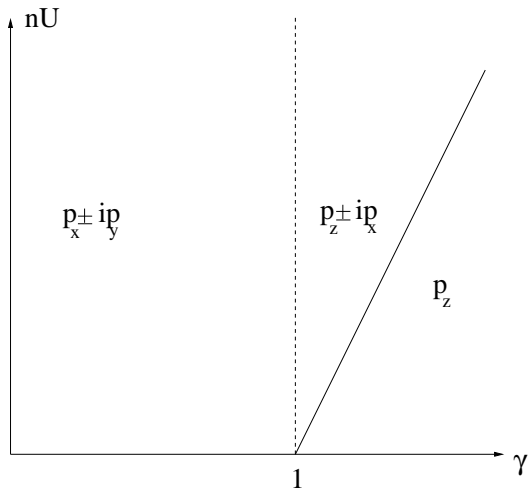}
\caption{Proposed phase diagram for a system of 
anisotropic lattice potentials, sketched as a function of the mean
interaction energy $nU$ and the  anisotropic
ratio $\gamma=\omega_{x,y}/\omega_z$.
}
\label{fig:phase}
\end{figure}

Let us consider the case of  cylindrical symmetry, i.e., 
$\omega_x=\omega_y=\gamma \omega_z$.
At $\gamma<1$, particles condense into the $p_x(Q_x)\pm ip_y (Q_y)$
state to minimize both the kinetic and interaction energy
as discussed in the main text.
The $\gamma=1$ point is subtle. 
Although the full Hamiltonian (because of the hopping term) 
does not possess the $SO(3)$ symmetry, the
condensation manifold recovers the $SO(3)$ symmetry at the mean field level.
The quantum fluctuation effect is 
expected to break this $SO(3)$ down to the
cubic lattice symmetry.

At $\gamma>1$, a quantum phase transition takes place 
from a time-reversal invariant polar condensate 
to a time-reversal symmetry broken TSOC state, as the boson density
increases (see Fig.~\ref{fig:phase}).
Let us parameterize the condensate as
\bea
\begin{pmatrix}
\avg{b_{z\Bk=\BQ_z}}\\
\avg{b_{x\Bk=\BQ_x}}
\end{pmatrix}
= {\sqrt n }
\begin{pmatrix}  \cos\chi \\ i\sin\chi
\end{pmatrix}\, 
\label{eq:<b>=SO2}
\eea
where $n$ is the boson density.
Then the mean field energy per unit volume is
\bea
E/{\cal V} &=& n (\omega_z\cos^2 \chi+ \omega_x\sin^2 \chi )
+\frac{U n^2}{2} (1-\frac{1}{3} \sin^2 2 \chi \nn \\
&=& n \bar \omega +\frac{U}{3} n^2 
+\Big\{ \frac{Un^2}{6} \cos^2 2 \chi -\frac{n}{2} \Delta \omega 
\cos 2\chi \Big \}, \nn \\
\eea
where $\bar\omega=\omega_z(1+\gamma)/2 $ and $\Delta \omega= (1-\gamma)
\omega_z$.
Minimizing the energy, we find
\bea
\cos 2 \chi=\left\{ 
\begin{array}{cc}
\frac{3}{2}\frac{\Delta\omega}{ n U} &
(n U > \frac{3}{2} \Delta\omega )
\\ 
1 & (n U < \frac{3}{2} \Delta\omega )
\end{array}
\right. .
\eea
When $n$ is small, kinetic energy dominates over  interaction;
Therefore a polar condensate in $p_z(Q_z)$ is favorable.
As $n$ increases larger, the interaction part becomes more important,
thus the TSOC state appears.
Again in the TSOC state, there is an $SO(2)$ symmetry (of $x$-$y$ plane
rotations) in the manifold of  the mean field
ground state, $p_z \pm i (\cos \alpha ~ p_x +\sin \alpha ~p_y)$ 
where $\alpha$ is an arbitrary azimuthal angle.
Fluctuation effects are expected to break this symmetry down to the 
tetragonal symmetry.

\bibliography{orbital_bec}

\end{document}